# Blink and you'll miss it - How Technological Acceleration Shrinks SETI's Narrow Detection Window.

## Michael A. Garrett[a,b,c] *

[a] *Jodrell Bank Centre for Astrophysics, Dept. of Physics & Astronomy, Alan Turing Building, Oxford Road, University of Manchester, M13 9PL, UK.* michael.garrett@manchester.ac.uk
[b] *Leiden Observatory, Leiden University, Postbus 9513, 2300 RA, Leiden, The Netherlands.*
[c] *Institute of Space Sciences and Astronomy, University of Malta.*

\* Corresponding Author

## Abstract

The search for extraterrestrial intelligence (SETI) has historically focused on detecting electromagnetic technosignatures, implicitly assuming that alien civilisations are biological and technologically analogous to ourselves. This paper challenges that paradigm, arguing that highly advanced, potentially post-biological civilisations may undergo rapid technological acceleration, quickly progressing beyond recognisable or detectable phases. We introduce a simple model showing that the technological acceleration rate ($\alpha$) of such civilisations can compress their detectable phase $\tau_d$ to mere decades, dramatically narrowing the temporal "detection window" in which their technosignatures overlap with our current capabilities. This framework offers a plausible resolution to the "Great Silence": advanced civilisations may be abundant and long-lived, but effectively invisible to present-day SETI methods. Consequently, our efforts must include but also evolve beyond the search for narrow-band communication signals in the radio and optical domains. Instead, we require an expanded, technology-agnostic strategy focused on persistent, large-scale manifestations of intelligence, such as broadband electromagnetic leakage, waste heat from megastructures, and multi-dimensional anomaly detection across extensive, multi-wavelength and multi-messenger datasets. Leveraging advanced artificial intelligence for unsupervised anomaly discovery, recursive algorithm optimisation, and predictive modelling will be essential to uncover the subtle, non-anthropocentric traces of advanced civilisations whose technosignatures lie beyond our current technological and cognitive frameworks.

## 1. Introduction

For many decades, the search for extraterrestrial intelligence (SETI) has relied on detecting familiar technosignatures that remain focused on detecting intentional narrow-band signals associated with powerful beacons in the radio and optical domains [1-13]. While these foundational methodologies have shaped the field, it seems not unrealistic to consider that highly advanced, potentially post-biological civilisations [14-16] might communicate or operate in ways fundamentally dissimilar to our past and indeed current technological paradigm [17-18].

It has long been recognised that the longevity of a technosignature is a critical factor in determining the likelihood of its detection (e.g. [19-22]). This paper further argues that successful detection also depends on the period during which our technologies overlap with those of the civilisations we seek. For advanced, post-biological civilisations, the acceleration of technological progress may be so rapid that this "detection window" is currently far narrower than previously assumed.

The concept of a limited window for communication with peer civilisations is not new. In 1973, Sagan introduced the "communication horizon" [23], postulating that a civilisation 1,000 years more advanced might be undetectable due to divergent technologies and a lack of interest in communicating. The work presented here advances this concept by proposing a model where the duration of this detection window is not fixed to an arbitrary value but is instead directly linked to the *rate* of technological acceleration, a factor that has become especially significant in the age of AI.

In this paper, we examine the implications of a narrow "detection window" for identifying advanced technological civilisations, particularly those that may be post-biological. We explore a range of accelerating technological factors that could influence both the production and detectability of technosignatures. Special attention is given to unconventional forms of technosignatures that might fall outside the traditional electromagnetic (EM) spectrum or manifest themselves as outliers in multi-wavelength and multi-messenger datasets. By leveraging emerging technologies, especially those driven by artificial intelligence, the aim of SETI must be to broaden the scope of our search and increase the chances of identifying civilisations that may be vastly different from our own.

The structure of this paper is as follows: Section 2 introduces a simple model showing how a narrow detection window $\tau_d$ emerges from the technological acceleration ($\alpha$) of a rapidly evolving civilisation, and how this constrains the likelihood of detecting their technosignatures with current instrumentation. Section 3 estimates some plausible values for $\tau_d$ and how this relates to the "Great Silence" [24]. Section 4 briefly explores a range of alternative technosignatures that may be produced by highly advanced civilisations, including those that may not rely on conventional electromagnetic emission. The future and critical role of AI in anomaly detection is also discussed. Finally, section 5 presents the main conclusions of the paper.

## 2. The narrow detection window for advanced, (post-biological) civilisations

Artificial intelligence (AI) has made extraordinary strides over the past decade, and especially in recent years [25]. This remarkable progress underscores how the timescales for technological advancement in AI are accelerated compared to the durations typical of Darwinian evolution [26]. Current trends suggest that AI performance is doubling every quarter, leading many to predict systems that could exceed human capabilities soon. This path from AGI (Artificial General Intelligence), which matches human cognitive abilities, to ASI (Artificial Superintelligence) which vastly surpasses them, could be extremely rapid. In particular, AGI may be a reality before end of this decade. There is, accordingly, growing speculation that the emergence of ASI may also occur much sooner than previously anticipated [27].

If humanity is able to steer and harness the development of this transformative technology, the consequences for scientific discovery and exploration could be profound. Even if control proves elusive [28-30], it is conceivable that a post-biological civilisation might emerge that continues to evolve and expand autonomously, liberated from its human origins and biological constraints [31]. Moreover, it is reasonable to assume that any existing extraterrestrial intelligences will have already undergone comparable transitions and progressed well beyond this phase [31-33]. Whether guided or emergent, AI is likely to play a pivotal role in greatly accelerating the technological trajectories of such advanced civilisations, driving them toward levels of capability far beyond anything presently conceivable within our own technological framework. Under such conditions, it is plausible to envision civilisations driven by ASI advancing their scientific understanding at unprecedented rates. As a consequence, the temporal window during which their technologies currently align with our own—and are therefore detectable—may be exceptionally brief.

The concept of a detection window, $\tau_d$, refers to the period during which a civilisation produces technosignatures that are detectable by another civilisation. Given the accelerating pace of technological development, particularly in the context of AI, it is plausible that advanced civilisations may transition rapidly from detectable to undetectable states. A civilisation becomes undetectable if it shifts to using technologies that we cannot detect. Of course, there may be other reasons that a civilisation can "go dark" e.g. transitioning to a post-biological state (transferring consciousness into machines or virtual environments) or by cloaking their emissions [34,35].

This section explores the mathematical implications of a narrow detection window on the probability of detecting technosignatures, and how this affects the traditional SETI paradigm.

**2.1 Detection Window model**

The central thesis of a narrow detection window can be formalised by considering the relationship between a civilisation's technological advancement and our capacity to detect it. The final term in the Drake equation, L, represents the communicative lifetime of a civilisation [36]. We propose a modification to this perspective, arguing that the critical variable is not the communicative lifetime of a civilisation, but the duration during which it produces technosignatures that are *detectable by us, now*.

Thus:

$$N_{det} = R_{\star}.f_p.n_e.f_l.f_i.f_c.\tau_d \qquad (1)$$

where $N_{det}$ the number of detectable civilisations, $R_{\star}$ is the rate of star formation averaged over the lifetime of the galaxy, $f_p$ is the fraction of stars with planets, $n_e$ is the mean number of planets in each planetary system with environments favourable for life, $f_l$ is the fraction of such planets that develop life, $f_i$ is the fraction developing intelligence, $f_c$ is the fraction that develop technology, and $\tau_d$ is the duration of detectability (or the window of detectability).

The first three astronomical terms of the equation are relatively well established ( $R_{\star}.f_p.n_e$ ) $\sim 0.1$ [23]) but the next three terms are not ($f_l.f_i.f_c$). Astronomers tend to assume highly optimistic values for these terms ($f_l.f_i.f_c$) $\sim 0.1$. while biologists suggest values that are much smaller [24]. Even if we adopt the optimistic values, we derive:

$$N_{det} \sim 0.01 \, \tau_d$$

For values of $\tau_d$ $\sim$ 100–200 years, N $\sim$ 1–2.

Before presenting the detection window model in more detail, it is important to consider the assumption of sustained exponential growth as we will use this going forward. Historically, the development of any single technology tends to follow a logistic (or "S") curve, with initial exponential growth that eventually slows and plateaus as it approaches physical or practical limits [37]. However, the overall technological capability of a civilisation can be viewed as a composite of many such overlapping curves. As one technology matures, a new paradigm-shifting technology often emerges, initiating a new phase of exponential growth.

From this perspective, continuous exponential advancement, as proposed by Kurzweil [38], may be a reasonable long-term approximation for the aggregate technological level of a civilisation, at least until fundamental physical limits are reached across all domains. In particular, Artificial Intelligence, represents a fundamental shift. Unlike previous revolutionary technologies that were domain-specific (e.g., steam power, electricity), AI is a foundational, general-purpose technology that acts as a universal catalyst for innovation itself. By optimising complex systems, discovering novel materials, and solving previously intractable problems across disciplines, AI has the potential to steepen the growth phase of existing technological curves and dramatically shorten the transition time to new ones. AI is best understood not as another technology subject to its own S-curve, but as a meta-technology that will boost innovation across all other domains. Its ability to process vast, multi-modal datasets and identify complex, non-obvious patterns will permit it to accelerate the research and development cycle itself. This effect becomes even more profound when considering the advent of AGI and ASI, which could introduce a recursive self-improvement loop that drives technological progress at a rate with no historical precedent.

This paper therefore proceeds with the exponential model of growth to explore its direct consequences, while acknowledging that plateaus or variable growth rates would extend the detection window for certain technological phases.

Let us then model a civilisation's technological level, K, as a continuous variable that increases over time, t. Our ability to detect technosignatures is confined to a specific range of this technological level, [ $K_{\min}$, $K_{\max}$]. Signatures produced by technologies below $K_{\min}$ may be too primitive or weak to be observed at interstellar distances, while those produced by technologies exceeding $K_{\max}$ may be based on physical principles or mediums currently unknown to us, rendering them invisible to our surveys.

The rate of technological advancement is the key factor. We model technological progress being made at an exponential rate [38]:

$$K(t) = K_0 e^{\alpha t} \qquad (3)$$

where $K_0$ is the initial technological level at the onset of growth (t=0), and $\alpha$ is the rate of technology acceleration. A larger value of $\alpha$ signifies a more rapid progression through technological paradigms.

For a pre-industrial civilisation in a "slow" developmental phase, $\alpha$ is much less than 1 ($\alpha \ll$ 1), resulting in essentially linear technological growth. Conversely, for a civilization on the brink of achieving recursively self-improving artificial intelligence, $\alpha$ would exceed 1 ($\alpha > 1$), triggering a period of rapid, exponential advancement.

Presently, $\alpha$ varies significantly across different technological domains. We therefore treat it as an aggregate measure, averaging progress across disciplines to estimate the timescale over which obsolete technologies are replaced or become obsolete.

Our detection window is defined by the time it takes for a civilisation's technology, $K(t)$, to traverse our detectable range, from $K_{\min}$ to $K_{\max}$. We can calculate $\tau_d$ the duration of detectability, by finding the time interval [ $t_1, t_2$] corresponding to the technological range [$K_{\min}, K_{\max}$].

$$K_{\max} = K_0 e^{\alpha t_2} \ \rightarrow t_2 = \frac{1}{\alpha} \ln\left(\frac{K_{\max}}{K_0}\right)$$

The duration of the detectable window is therefore:

$$\tau_d = t_2 - t_1 = \frac{1}{\alpha}\left[\ln\left(\frac{K_{\max}}{K_0}\right) - \ln\left(\frac{K_{\min}}{K_0}\right)\right]$$

This simplifies to:

$$\tau_d = \frac{1}{\alpha}\ln\left(\frac{K_{\max}}{K_{\min}}\right) \qquad (4)$$

Equation (4) demonstrates that $\tau_d$ is inversely proportional to $\alpha$, but only logarithmically dependent on the ratio $K_{\max}/K_{\min}$. As a civilisation's technological acceleration $\alpha$ increases, particularly under the influence of AI-driven growth, the time it remains detectable for any given class of technosignatures may shrink significantly.

We note that this model assumes a single acceleration rate, $\alpha$, for a given civilisation. In reality one would expect a wide distribution of $\alpha$ values across the galaxy, influenced by factors such as biology, culture, AI architecture choices, or even deliberate societal decisions to limit AI development. A consequence of this would be a selection effect: SETI searches would be inherently biased toward detecting civilisations with lower values of $\alpha$, as their longer detection windows $\tau_d$ would make them statistically easier to find. Our search may therefore preferentially uncover the "slow growers" rather than the most rapidly advancing civilisations.

## 3. Estimating Plausible Values for $\tau_d$

To explore this model further, we can estimate plausible values for the parameters based on past and current human experience. The critical variable is $\alpha$, the average rate of technological acceleration. The value of $\alpha$ will have the most significant impact on detectability. The units of $\alpha$ are inverse time, and can be considered in years, centuries, etc. and it directly relates to the technology "doubling time" via the relationship:

$$\alpha = \frac{\ln(2)}{\text{doubling time}} \quad (5)$$

To ground our estimates, we can refer to observed and current technological growth trends on Earth. Fuller introduced the concept of the Knowledge Doubling Curve [39], estimating that up until ~1900, human knowledge doubled approximately every 100 years and by ~ 1950, the doubling time had shortened to ~ 25 years.

Based on these trends, we adopt three regimes of technological growth:

• **Slow Growth:** A technology doubling time of 100 years corresponds to α=ln(2)/100 ≈ 0.007.

This represents a conservative (pre-industrial) trajectory.

• **Moderate Growth:** A doubling time of 25 years, reflecting technological progress in the mid-to-late 20th century, yields α=ln(2)/25 ≈ 0.03.

This is characteristic of a steadily advancing, industrialised society.

• **Rapid Growth:** A doubling time of 5 years – similar to the kind of progress we currently see in AI models (doubling in capability on timescales of ~ 1 year or less) and Moore's Law [40] (a doubling of computing processing power approximately every 2 years). While this growth is confined to a particular class of technologies, AI assisted design and its unique discovery potential is ushering in a new era of AI driven growth with much larger values of α likely to be realised across a wide-range of different technologies, including those that are today considered to be stagnant.

A doubling time of 5 year seems reasonable for our purposes, and corresponds to an acceleration rate of α=ln(2)/5 ≈ 0.14.

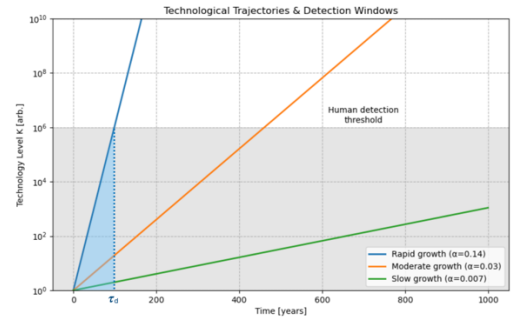

*Figure 1: A plot of a civilisation's Technical Level (K) versus time. The window of detection is defined as the area under the growth lines that intersect with the human detection threshold. This is illustrated for the "rapid growth" curve (blue).*

The ratio $K_{max}/K_{min}$ represents the total improvement in a given technological paradigm from the point it becomes detectable from interstellar space ($K_{min}$) to the point it is no longer detectable ($K_{max}$). A factor of $10^6$ seems a reasonable estimate for the gains made in a technology before it becomes obsolete. In this case, ratio $ln(K_{max}/K_{min}) = \ln(10^6) \approx 14$.

Assuming $K_{max}/K_{min} \sim 10^6$, we can estimate $\tau_d$:

- For slow growth (α ≈ 0.007): $\tau_d$ = 14/0.007 ≈ 2000 years.
- For moderate growth (α ≈ 0.03): $\tau_d$ = 14/0.03 ≈ 500 years.
- For rapid growth (α ≈ 0.14): $\tau_d$= 14/0.14 ≈ 100 years.

Figure 1 plots the detection window $\tau_d$ for the various values of α. We note that choosing much larger values for $K_{max}/K_{min}$ e.g. $10^{10}$ or $10^{15}$ only changes values of $\tau_d$ by a factor of a few. This can be seen in Figure 2 which plots $\tau_d$ versus α for a range of $K_{max}/K_{min}$.

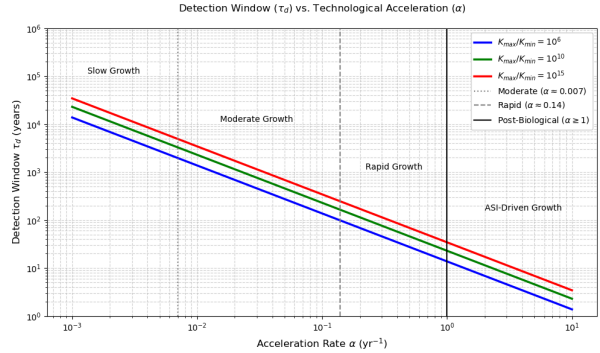

*Figure 2: τ_d plotted against α for a range of $K_{max}/K_{min}$. The results are very weakly influenced by the value chosen for this ratio.*

### 3.1 Acceleration in Post-Biological Civilisations

The landscape of technological advancement shifts dramatically when considering post-biological civilisations driven by ASI that can recursively and autonomously improve [41]. Freed from biological constraints, such entities could enhance their capabilities rapidly, constrained only by resources and the laws of physics [42]. For instance, should a post-biological civilisation experience a technological doubling time of 1 year or less, the acceleration rate becomes α > 0.7. Such rapid advancement implies that the window for their technosignatures to be detectable with our current technologies could shorten to a few decades.

### 3.2 Discussion

The simple framework presented in this paper suggests that the detectability of advanced civilisations is fundamentally constrained by the temporal alignment between their technosignature production and our observational capabilities. Our analysis demonstrates that even long-lived civilisations may produce observable signatures for only fleeting periods, particularly if they undergo post-biological transitions characterised by recursive self-improvement (α ≫ 0.7).

The inverse relationship between α and $\tau_d$ (equation 4) implies that rapidly advancing civilisations inevitably outpace our current ability to detect them. For post-biological civilisations with α ≥ 1 yr⁻¹, detectable windows shrink to $\tau_d \lesssim 20$ years, a timescale cosmologically insignificant compared to the ~100,000-year span of human civilisation, let alone the billion-year habitability of Earth-like planets. With detection windows much shorter than the typical values of L assumed in the Drake equation (equation 1), the probability of overlap between our current search capabilities and the technological phase of an advanced or post-biological civilisation will be minimal.

The model primarily ties $\tau_d$ to the active technological phase of a civilisation. This is naturally somewhat simplistic - several factors could complicate this picture and potentially widen the effective detection window. Economic inertia and the vast infrastructural costs of deploying new technologies or the construction of astro-engineering mega-structures, could slow the pace at which a society abandons older, detectable methods [43]. Socio-political factors, such as a moratorium on certain technologies for safety reasons or natural disasters that require a technological restart, could also introduce plateaus in development. Moreover, a civilisation might create "legacy technosignatures" such as powerful, long-lived beacons that are deliberately designed to operate autonomously for millennia, long after the creators have moved to other communication modes. The lifespan of the transmitter itself could then become the dominant factor in terms of detectability. Finally, there are fundamental physical constraints; for example, unrestrained growth in energy consumption is ultimately limited by thermodynamic considerations, such as planetary waste heat, which could impose a ceiling on certain expansionist trajectories [44].

Critically, this model aligns with humanity's own recent technological trajectory. For example, our radio emissions have transitioned from a few fixed, high-power, narrow-band omnidirectional, low-frequency (< 1 GHz) broadcasting antennas [45], to billions of mobile, low-power, broadband, highly directional, high frequency digital communication systems. This has all happened well within 50 years [46]. If this sort of trajectory is universal, the galaxy could host myriad civilisations whose electromagnetic signatures were briefly visible before being replaced by something else, something better.

This dynamical perspective offers a compelling resolution to the 'Great Silence,' reinforcing Martin

Rees's observation that "absence of evidence is not evidence of absence" (as quoted in [47]). The lack of detectable technosignatures may reflect not a scarcity of civilisations but the brevity of their detectable phases within any given technological paradigm. Rather than pointing to rarity or self-destruction, this "transcendence filter" suggests that civilisations evolve beyond our current observational thresholds. A civilisation might endure for eons, yet the period during which it uses recognisable technologies, such as high-power, narrow-band, omnidirectional radio transmissions, could be vanishingly small. Consequently, the probability of our own search efforts coinciding with this brief window is tiny, reframing the silence not as evidence of absence, but as evidence of extreme technological disparity.

## 4. Implications for SETI Observing Strategies

The concept of a narrow detection window, as discussed in previous sections, suggests that advanced, potentially post-biological civilisations may produce technosignatures that are detectable for only brief periods of time. This necessitates a re-evaluation of current SETI strategies to increase the likelihood of detecting such civilisations.

Future technosignature searches should consider the prioritisation technology-agnostic approaches that focus on macro-scale manifestations of advanced activity, such as large-scale engineering or energy harvesting. They should target persistent signatures that are more likely to remain observable over long timescales, even as civilisations undergo profound technological transformations. These searches should also preferably be statistically robust, leveraging systematic multi-wavelength/multi-messenger wide-field or all-sky surveys and employing AI-driven techniques (see also section 4.4) to exhaustively detect, classify, and prioritise non-natural anomalies. Such strategies would greatly enhance our ability to detect civilisations whose observable imprints are subtle, rare, or dispersed across vast multi-dimensional datasets.

### 4.1 Technology-agnostic & persistent approaches

Technology-agnostic searches target macro-scale anomalies, such as megastructures or large-scale engineering, without assuming specific technologies. These signatures, rooted in fundamental physical principles (e.g., energy conservation), are detectable regardless of implementation. Megastructures like Dyson spheres or swarms, for instance [48], could produce observable infrared excesses [49-53] or anomalous stellar dimming [54], and are likely to persist for millions of years as enduring features of a civilisation's energy infrastructure.

Wide-field surveys such as those by the Vera C. Rubin Observatory [55] and Square Kilometre Array (SKA) [56] will enable systematic searches for such rare anomalies across stellar and galactic populations. By analysing large datasets for unusual light curves, spectral energy distributions, unusual object patterns/motions or astrometric deviations, astronomers can statistically distinguish candidate technosignatures from natural phenomena. Promising anomalies can then be followed up with deep, multiwavelength observations, including targets within our own Solar System [57]. This approach complements traditional SETI by focusing on the physical consequences of advanced civilisations in addition to their communication methods, thereby broadening the range of persistent technosignatures we are sensitive to.

### 4.2 Searching for Broadband Emission Across the Electromagnetic Spectrum

A comprehensive survey of the electromagnetic spectrum is essential for detecting both transient and persistent technosignatures. Even within the radio domain, large regions of the spectrum remain unexplored. Most SETI efforts still target narrowband signals in the "water hole" (1–3 GHz), despite the shift in terrestrial technologies toward low-power, broadband systems at higher frequencies—a trend that is likely to continue [46].

Recent work has extended searches beyond 10 GHz [58], with exploratory efforts reaching into the millimeter regime [59]. At the other extreme, sub-10 MHz frequencies are attracting interest, though meaningful studies require instruments above the ionosphere—ideally in space or on the lunar farside [60].

Narrowband signals remain a principal focus of SETI, as they are exceedingly rare in nature and therefore constitute strong candidates for artificial origin, although their detection is complicated by pervasive terrestrial interference. Conversely, it is also plausible that technological advancement would not lead to the abandonment of radio beacons, but to their enhancement. The same principles of signal-to-noise and energy efficiency that make narrowband signals appealing for initial detection remain valid regardless of technological level. An advanced AI could design and operate extremely powerful, efficient, and targeted phased-array transmitters for interstellar or even intergalactic communication, using them to deliberately signal its presence. In this scenario, technological growth would *widen* or sustain the detection window for such intentional beacons, rather than closing it.

Broadband signals [61] present greater challenges for discrimination against astrophysical backgrounds; however, simulations indicate that the aggregate leakage from Earth's mobile communication systems and civilian and military radar would constitute a detectable broadband source to an advanced civilisation [46]. Detecting comparable low-power emissions from extraterrestrial sources may necessitate the use of Very Long Baseline Interferometry, which can both spatially resolve a civilisation's emission from its host star and function as a high-brightness temperature filter against the distributed and diffuse radio sky [62]. Such techniques are expected to play an increasingly significant role in future broadband SETI efforts.

### 4.3 Beyond Electromagnetism: A Multi-Messenger Approach

While electromagnetic searches remain foundational to the field, a sufficiently advanced civilisation might transcend electromagnetic-based technologies, employing information carriers that are more secure, less subject to diffraction limitations, or less susceptible to natural background noise. In particular, the era of multi-messenger astrophysics [63] opens entirely new observational avenues for the discovery of novel technosignatures, potentially involving neutrinos, quantum information carriers, high energy particles, dark matter/energy and gravitational waves [64]. Of these possibilities, searching for anomalous gravitational-wave signatures might be most within our current reach [65]. While the artificial generation of gravitational waves would demand energy far exceeding current human capabilities, a Kardashev Type II or III civilisation could plausibly achieve this by manipulating compact binary mergers, accelerating megastructures to extreme velocities, or constructing massive, highly deformed rotators. The latter would be especially significant, as it would produce a continuous gravitational-wave signal, analogous to deliberate high-power radio beacons. Gravitational waves (GWs) also possess several compelling properties as potential probes of technosignatures: (i) they propagate through space-time at the speed of light without being absorbed or scattered by intervening matter or electromagnetic fields, enabling transmission across cosmological distances [66, 67]; (ii) while their power follows the same inverse-square law as electromagnetic radiation, the measured spatial strain decreases as $1/r$ [69]; (iii) artificially generated GWs would likely produce continuous and/or distinct, non-astrophysical waveforms, clearly differentiating them from the characteristic "chirps" of natural sources [69]; (iv) GW observatories are inherently sensitive to the entire sky; and (v) the field is on the cusp of dramatic growth, with next-generation facilities [70-72] poised to achieve sensitivities several orders of magnitude better than current instruments and operating across a wider range of frequencies. However, GW-based communication faces significant challenges, including the immense energy required for generation and the low bandwidth of plausible sources (typically < 1 kHz) would severely limit data rates. Neutrinos may be a more practical alternative – they also penetrate unimpeded through interstellar matter largely unimpeded, their production in a directional beam is energetically less demanding, and their particle nature allows information to be encoded by modulating the beam's intensity, enabling much higher data rates [73].

Similarly, while the manipulation of dark matter or dark energy may appear highly speculative, it cannot be excluded that a post-biological civilisation could develop such capabilities. Given that dark matter constitutes roughly 85 % of the universe's matter content [74], advanced civilisations that can detect it may also discover ways to harness or modify it for communication or energy production. Likewise, the local manipulation of dark energy could generate observable space-time anomalies. Although our current understanding of these components remains very limited, future progress in cosmology and particle physics may open pathways for identifying such exotic technosignatures.

In the future, microlensing, using either the solar gravitational lens or other stellar lenses, should be exploited to enhance the sensitivity of multiwavelength and multi-messenger instrumentation by several orders of magnitude [75]. Although the range of targets is naturally rather limited, the amplification that mircrolensing enables, effectively extends the detection window $\tau_d$ for both electromagnetic and multi-messenger technosignatures.

## 4.4. Leveraging Artificial Intelligence

Machine learning has already proven its value to technosignature research [76-81], and the sheer scale of future SETI datasets, spanning the electromagnetic spectrum and, ideally, extending into multi-messenger domains. These rich and complex data sources will far exceed the capacity of traditional human-driven analysis. Artificial intelligence offers clear solutions to many of these challenges: processing unprecedented data volumes, distinguishing subtle patterns and anomalies, developing online adaptive observing strategies, and enabling real-time signal processing and RFI mitigation.

Yet AI's role extends beyond enhancing data analysis or scaling anomaly detection. It will introduce a fundamental shift in how we conceptualise and conduct the search for technosignatures. This includes multi-dimensional anomaly detection across electromagnetic and multi-messenger domains, the construction of highly accurate models of the "natural" universe, automated rapid-response follow-up observations, and the integration of cross-disciplinary insights. Perhaps most transformative is the development of self-improving search algorithms and decision-making frameworks that operate free from anthropocentric assumptions [82], expanding SETI's reach into previously unimagined discovery spaces.

Equally transformative, though less immediately apparent, is ASIs potential to revitalise under-explored areas of SETI research. Natural language processing, powered by advanced AI [83], will better prepare us to decode potential alien communications or interpret the semantic content of detected signals. Furthermore, ASI-driven simulations of humanity's own technosignatures, as well as those of hypothetical advanced or post-biological civilisations, could predict optimal detection windows and better guide search strategies. By modeling the evolution of technosignatures across different technological paradigms, ASI can help identify the most likely signatures of civilisations at various stages of development. Ultimately, this approach can expand our ability to recognise technosignatures that fall outside traditional assumptions about communication or technology, broadening the scope of what we consider detectable.

Ultimately, the concept of a narrow detection window necessitates a paradigm shift in SETI strategies. By trying to adopt technology-agnostic approaches, expanding our search across the electromagnetic spectrum and into multi-messenger domains, and leveraging the power of artificial intelligence, we can significantly enhance our ability to detect advanced civilisations. However, it is crucial to acknowledge that even these expanded strategies may face limitations in detecting the most advanced, post-biological societies that may have chosen to minimise their observable footprints [20, 21]. As we continue to refine our search methods, we must remain open to the possibility that these signatures may be subtle, rare, or fundamentally different from our current expectations. This underscores the importance of continued innovation in SETI methodologies and the need for interdisciplinary collaboration to push the boundaries of our current search capabilities.

## 5. Conclusions

The search for extraterrestrial intelligence has, for decades, been shaped by anthropocentric assumptions about how other civilisations might communicate and what their technosignatures might look like. Such assumptions can be very narrow, particularly when considering advanced, potentially post-biological civilisations undergoing rapid technological transitions. This paper has introduced the concept of a narrow detection window $\tau_d$, the brief period during which a civilisation produces technosignatures that remain detectable by our current instrumentation. As the rate of technological acceleration, $\alpha$, increases, this window may contract to mere decades, reframing the "Great Silence" [24] as a symptom of extreme technological disparity.

While this paper's model adopts a simple exponential growth trajectory to explore the consequences of rapid advancement, the path of a civilisation is likely a more complex interplay of successive technological paradigms, moderated by socio-economic inertia, political considerations, and fundamental physical limits. Furthermore, some advanced societies may choose to enhance traditional beacons rather than replace them, and long-lived "legacy technosignatures" cannot be ruled out although we might expect to have already detected them. However, even when accounting for these moderating factors, the central thesis remains robust: the alignment of technological capability between a searching and a developing civilisation is likely to be brief and rare. This may be particularly the case for post-biological civilisations where the limitations are partially relaxed by constraints governed primarily by resources and the fundamental laws of physics.

If this interpretation is broadly correct, the implications for SETI are profound. The traditional search for narrowband electromagnetic signals should be complemented by strategies that prioritise technology-agnostic, statistically robust approaches, focusing on persistent macroscale anomalies, such as megastructures or other manifestations of largescale astro-engineering, including broadband leakage. The future of SETI surely lies in an anomaly-based approach that is based on a deep understanding of astrophysics and the natural universe - searching for persistent, large-scale physical consequences of astro-engineering, in addition to intentional messages or beacons. This will rely on leveraging the full observational spectrum, from radio and infrared to gravitational waves and exotic high energy particles. Our greatest ally in this profound challenge will be our own developing AI, deployed to find the faint fingerprints of its far more advanced counterparts. AI is poised to revolutionise technosignature searches through real-time data triage, sophisticated anomaly detection, multi-dimensional data integration, and self-improving development unconstrained by human preconceptions.

In conclusion, the search for extraterrestrial intelligence is also a search for our own future. As humanity itself approaches potential post-biological transitions, SETI's greatest challenge, and opportunity, lies in recognising that the universe's most advanced civilisations may not conform to our expectations. By embracing this uncertainty, we open the door to discoveries that could redefine not only our place in the cosmos but our understanding of the breadth of intelligence itself. Critically, this expanded and open approach does not discard traditional SETI but complements it, acknowledging that while some civilisations might still emit recognisable electromagnetic signals, others may operate in ways we cannot yet imagine. This exponential growth in technological advancement should significantly enhance our own chances of detecting extraterrestrial intelligence. However, this optimism is somewhat tempered by the uncertain trajectory of human intelligence in the face of rapidly evolving AI. One sobering question is whether human intelligence will still be around to witness all of this. In the end, we may need to be resigned to the possibility that it will be our intelligent machines, searching for theirs.

### Declaration of competing interest

The author declares that he has no known competing financial interests or personal relationships that could have appeared to influence the work reported in this paper.

### Acknowledgements

This research did not receive any specific grant from funding agencies in the public, commercial, or not-for-profit sectors.